%% file: eprint_szczurek.tex
\documentclass[12pt]{article}
\usepackage{graphicx}

\def\Krakow{Institute of Nuclear Physics, Polish Academy of Sciences,
  PL-31-342 Krak\'ow, Poland}
\def\Rzeszow{Faculty of Mathematics and Natural Sciences, University of
  PL-35-310 Rzesz\'ow, Poland}
\def\support{\footnote{Work supported by the Polish National Science
    Center grant DEC-2014/15/B/ST2/02528.}}

\def\Title#1{\begin{center} {\Large #1 } \end{center}}
\def\Author#1{\begin{center}{ \sc #1} \end{center}}
\def\Address#1{\begin{center}{ \it #1} \end{center}}

\newenvironment{Abstract}{\begin{quotation}  }{\end{quotation}}
\newenvironment{Presented}{\begin{quotation} \begin{center} 
             PRESENTED AT\end{center}\bigskip 
      \begin{center}\begin{large}}{\end{large}\end{center} \end{quotation}}
\def\Acknowledgements{\bigskip  \bigskip \begin{center} \begin{large}
             \bf ACKNOWLEDGEMENTS \end{large}\end{center}}

\input econfmacros.tex

\begin{document}
\begin{titlepage}

\vfill
\Title{Production of two $J/\psi$ mesons in proton-proton
collisions \\ at the LHC}
\vfill
\Author{Antoni Szczurek\support}
\Address{\Krakow}
\Address{\Rzeszow}
\Author{Anna Cisek}
\Address{\Rzeszow}
\Author{Wolfgang Sch\"afer}
\Address{\Krakow}

\vfill
\begin{Abstract}
We discuss inclusive production of pairs of $J/\psi$ quarkonia.
Both leading-order single parton (box) contribution
and double-parton contributions are included.
In addition we discuss a contribution of higher-order two-gluon exchange
and a feed down from double $\chi_c$ production.
The second two contributions are important for large rapidity
distance between the two $J/\psi$ mesons and for relatively large
cut-offs on quarkonia transverse momenta.
A relation to current experiments is discussed.
\end{Abstract}
\vfill
\begin{Presented}
EDS Blois 2017, Prague, \\ Czech Republic, June 26-30, 2017
\end{Presented}
\vfill
\end{titlepage}
\def\thefootnote{\fnsymbol{footnote}}
\setcounter{footnote}{0}
%

\section{Introduction}

The production of single $J/\psi$ mesons in proton-proton collsions was 
a vivid topic for more than three decades \cite{review}.
Several differential distributions were measured.
The nonrelativistic perturbative QCD is the standard theoretical 
approach in this context.
There is no agreement and comonly accepted explanation of the many
experimental data collected for years.
 
Recently there are several experimental results also for double $J/\psi$
production
\cite{D0_jpsijpsi,LHCb_jpsijpsi_7TeV,ATLAS_jpsijpsi,CMS_jpsijpsi,
LHCb_jpsijpsi_13TeV}. Also here there are puzzles that require
dedicated studies.
The leading-order result of the order of ${\cal O}(\alpha_s^4)$
is well known for some time \cite{BSZ2011}.
Also double-parton mechanism was discussed in this context.

Here we discuss several mechanisms relevant for understanding
current measurements. In addition to the leading-order contribution we
include also double-parton scattering mechanism.
The normalization parameter of the corresponding cross section, 
the so-called $\sigma_{\rm eff}$, is unknown. 
$\sigma_{\rm eff} \approx 15 \, \rm{mb}$ was obtained for other double parton 
scattering processes \cite{Astalos}.
The $\sigma_{\rm eff}$ parameter adjusted to some of the recent
double $J/\psi$ measurements were found to be much lower.
How this result can be understood?

Here we discuss two more mechanisms of double $J/\psi$ production
not included in theoretical analyses. We will discuss their main
characteristics and possible consequence for the value of $\sigma_{\rm eff}$
extracted from the experimental studies.

\section{Discussed mechanisms}

\begin{figure}[htb]
\centering
\includegraphics[width=4cm]{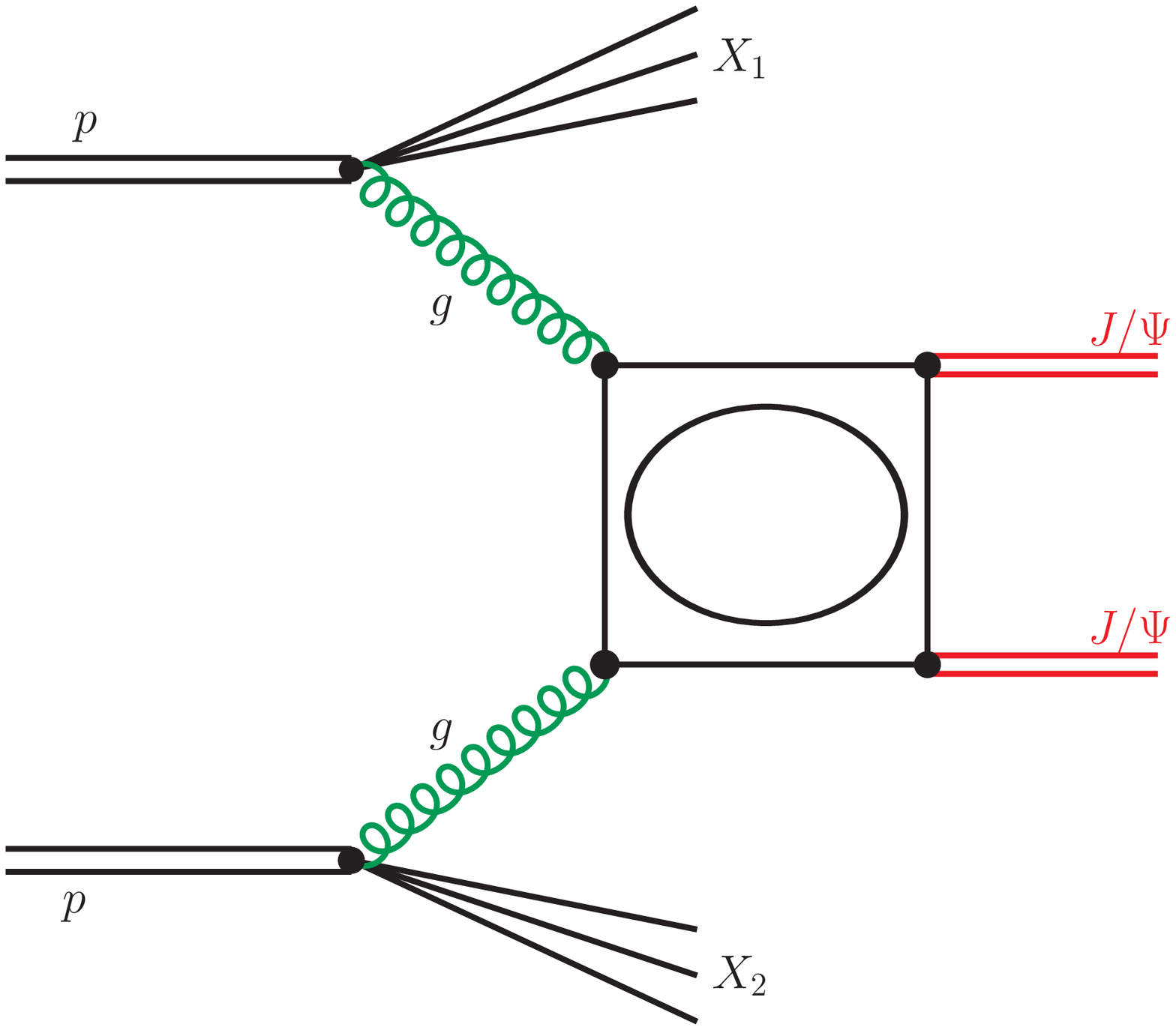}
\includegraphics[width=4cm]{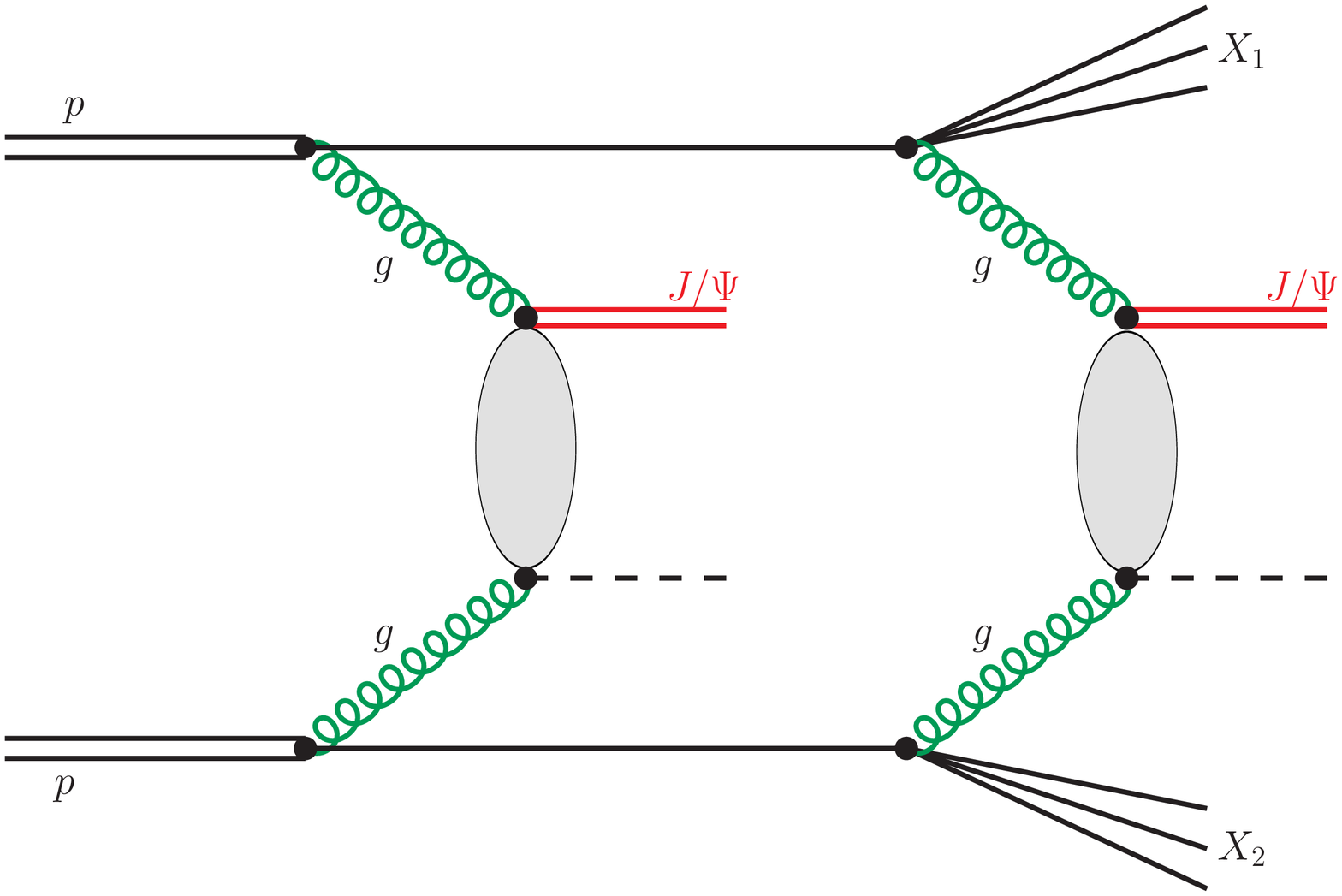}\\
\includegraphics[width=4cm]{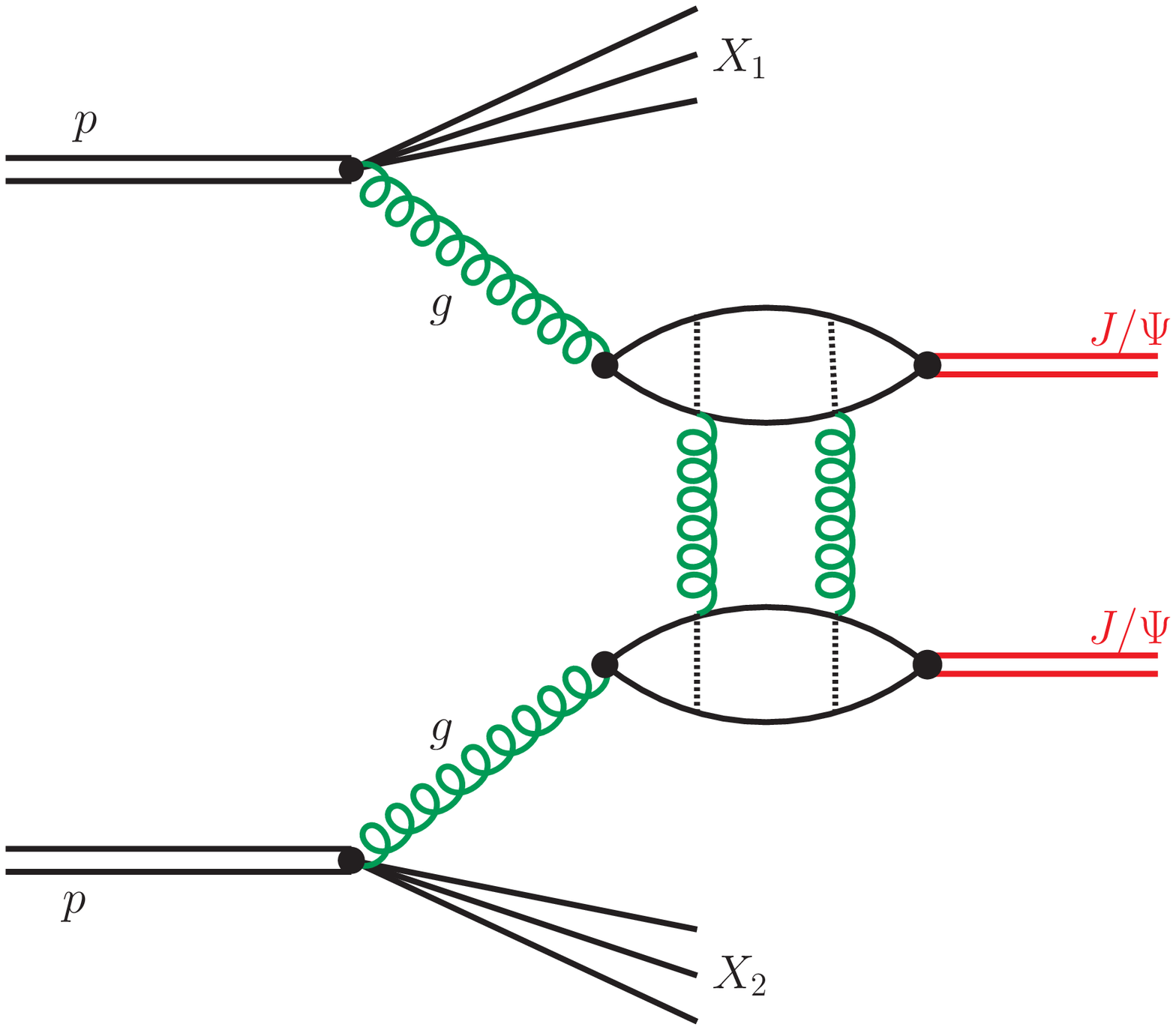}
\includegraphics[width=4cm]{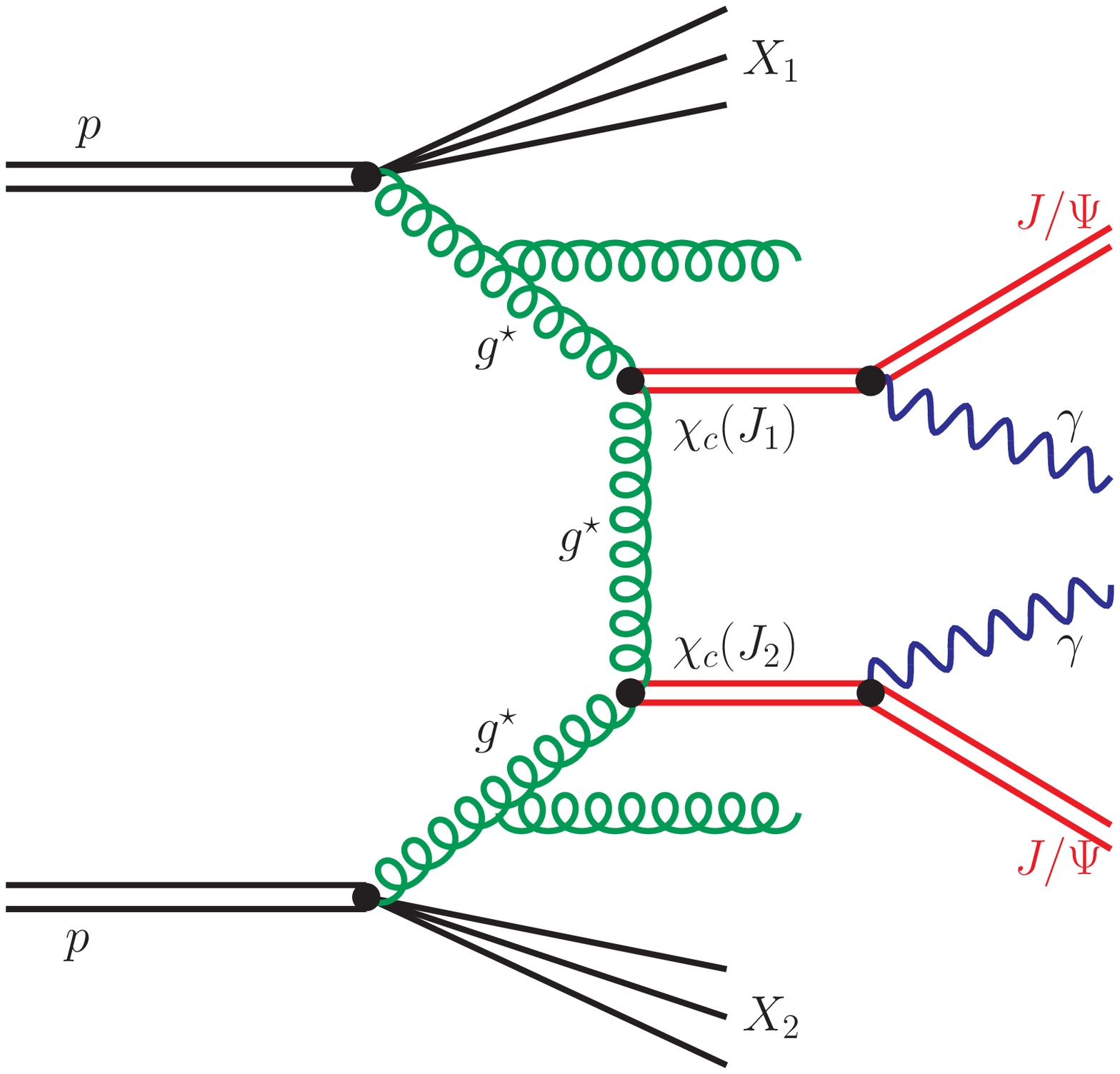}
\caption{The mechanisms considered in our analysis.}
\label{fig:mechanisms}
\end{figure}

The mechanisms considered in our analysis are shown in 
Fig.\ref{fig:mechanisms}.

The leading-order contribution is represented here in a bit
symbolic representation (the inserted circle). 
The circle represent extra gluon.
Some explicit Feynman diagrams were presented e.g.in our previous paper
\cite{BSZSS2013}.

The double-scattering diagram is presented here also in a bit generic
form. The blobs shown in the figure represent several mechanisms
of single $J/\psi$ production discussed e.g. by A. Cisek \cite{C2017} 
at this workshop.

The two-gluon exchange mechanisms was already considered in our previous
paper \cite{BSZSS2013}. It was shown there that it contributes to the
region of large rapidity distances between the two produced $J/\psi$ 
mesons.

The last mechanism was not considered in regular papers,
but was already discussed recently in some conference talks
by the present speaker. Here we only sketch main features of the
mechanism. The details will be presented elsewhere \cite{CSS2017}.
 
The contribution of the first mechanism is calculated here
in the $k_t$ factorization approach. 
In the $k_t$-factorization approach the corresponding differential cross
section can be written as:
\begin{eqnarray}
&&\frac{d \sigma(p p \to J/\psi J/\psi X)}
{d y_{1} d y_{2} d^2 p_{1t} d^2 p_{2t}}
 = 
\frac{1}{16 \pi^2 (x_1 x_2 s)^2} \int \frac{d^2 q_{1t}}{\pi} \frac{d^2 q_{2t}}{\pi} 
\overline{|{\cal M}_{g^{*} g^{*} \rightarrow J/\psi J/\psi}^{\rm off-shell}|^2} 
\nonumber \\
&& \times \;\; 
\delta^2 \left( \vec{q}_{1t} + \vec{q}_{2t} - \vec{p}_{1t} - \vec{p}_{2t} \right)
{\cal F}_g(x_1,q_{1t}^2,\mu_{F}^{2}) 
{\cal F}_g(x_2,q_{2t}^2,\mu_{F}^{2}) \; .
\label{kt_fact_gg_jpsijpsi}
\end{eqnarray}
The relevant off-shell matrix elements were obtained first in 
\cite{JJ_kt}.
The Kimber-Martin-Ryskin unintegrated distributions were used
in practical calculations.

In the present analysis we assume a simple factorized ansatz
for the double scattering cross section.
The the differential cross section for double $J/\psi$ production
can be writen as:
\begin{equation}
\frac{d \sigma(pp \to J/\psi J/\psi)}{d y_1 d^2 p_{1t} d y_2 d^2 p_{2t}} =
\frac{1}{2 \sigma_{\rm eff}} \cdot
  \frac{d \sigma(p p \to J/\psi X)}{d y_1 d^{2}p_{1t}} \cdot
  \frac{d \sigma(p p \to J/\psi X)}{d y_2 d^{2}p_{2t}} \; .
\end{equation}
$\sigma_{\rm eff}$ is responsible for the overlap of partonic
densities of colliding protons. It is often treated as a free parameter
to be fitted to experimental data. 
How to use single $J/\psi$ cross section is to some extent an open
issue. One possibility is to calculate it in some approach
($k_t$-factorization for instance). Another possibility is to
parametrize experimental data. Here we follow \cite{KKS2011,BK2016}
and write cross section in terms of a ficticious matrix element
with free parameters fitted to experimental data.
Quite nice fits to experimental data can be obtained.
This will be shown elsewhere \cite{CSSB2018}.

The double $\chi_c$ contribution is calculated in the
$k_t$-factorization approach with a formula similar as that for
the double $J/\psi$ production.
The matrix element for the $g^* g^* \to \chi_c(J_i) \chi_c(J_j)$
subprocess can be written in a somewhat simplified (omitting spins
of $\chi_c$ mesons) way as
\begin{eqnarray}
{\cal M}^{ab}_{\mu \nu} &=& 
V^{ac}_{\mu \alpha}(q_1,p_1-q_1)
{-g^{\alpha \beta} \delta_{cd} \over \hat t} 
V^{db}_{\beta \nu}(p_2-q_2,q_2) \nonumber \\
&+& 
V^{ac}_{\mu \alpha}(q_1,p_2-q_1)
{-g^{\alpha \beta} \delta_{cd} \over \hat u}
V^{db}_{\beta \nu}(p_1-q_2,q_1) \; .
\label{chicchic_amplitude}
\end{eqnarray}
To obtain the $k_t$-factorization amplitude one should contract the
above tensorial amplitude with the polarization vectors of the 
off-shell gluons \cite{CSS2017}.

There are several combinations of the $\chi_c(J_i) \chi_c(J_j)$
final states:
\begin{eqnarray}
(a) &&\chi_c(0) \chi_c(0) , \nonumber \\
(b) &&\chi_c(0) \chi_c(1) , \nonumber \\
(c) &&\chi_c(0) \chi_c(2) , \nonumber \\
(d) &&\chi_c(1) \chi_c(1) , \nonumber \\
(e) &&\chi_c(1) \chi_c(2) , \nonumber \\
(f) &&\chi_c(2) \chi_c(2) . \nonumber
\end{eqnarray}
The branching fractions in the $\chi_c \to J/\psi+\gamma$
decays \cite{PDG} cause that only (d),(e) and (f) cases are of practical
meaning.

More details will be discussed in our original papers \cite{CSS2017,CSSB2018}.

\section{Selected results}

\begin{figure}[htb]
\centering
\includegraphics[width=8cm]{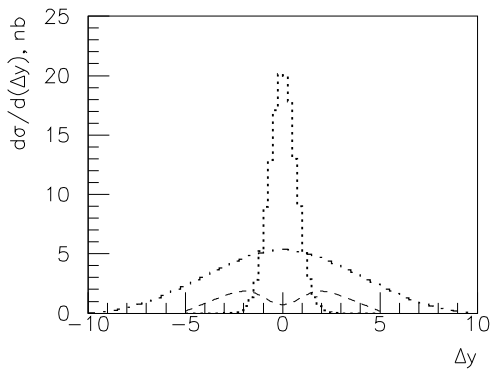}
\caption{Distribution in rapidity difference of the two
$J/\psi$ mesons for the LHCb kinematics.}
\label{fig:dsig_dydiff_LHCb}
\end{figure}

Here we wish to show only some selected results.
In Fig.\ref{fig:dsig_dydiff_LHCb} we show result
obtained in \cite{BSZSS2013} for the LHCb cuts.
There leading order, DPS and two-gluon exchange contributions were
considered. Clearly for the LHCb kinematics the leading order (box)
contribution dominates. However, when going to large rapidity distances
between the two $J/\psi$ mesons one has to include also
the DPS contribution (dash-dotted line) and even two-gluon exchange
contribution (dashed line) multiplied by an extra factor \cite{BSZSS2013}.

\begin{figure}[htb]
\centering
\includegraphics[width=8cm]{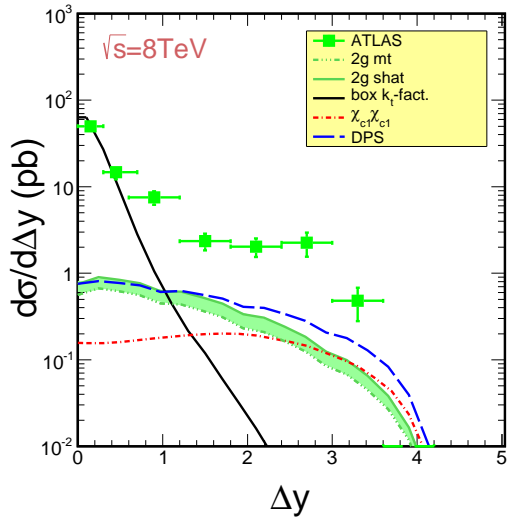}
\caption{Distribution in rapidity difference of the two
$J/\Psi$ mesons for the ATLAS kinematics.}
\label{fig:dsig_dydiff_ATLAS}
\end{figure}

Recently we have considered also the situation relevant for 
a recent ATLAS experiment \cite{ATLAS_jpsijpsi}.
Here -2.1 $< y <$ 2.1 and $p_t >$ 8.5 GeV (and some other cuts on
muons) conditions are imposed experimentally. 
The situation at the large transverse momenta 
($p_t >$ 8.5 GeV) is slightly different than for the LHCb experiment where
relatively low transverse momenta were measured.
In Fig.\ref{fig:dsig_dydiff_ATLAS} we show distribution in 
the rapidity difference for the ATLAS kinematics.
We show all four discussed here contributions.
In this preliminary calculation only one combination of 
$\chi_c(J_i) \chi_c(J_j)$ for $J_i$ = 1 and $J_j$ = 1 was included.
All combinations will be shown in Ref.\cite{CSSB2018}.
As for the LHCb experiment the dominant box contribution is concentrated
at low rapidity distances. Both two-gluon exchange
and double-$\chi_c$ contributions have a similar shape 
as the double scattering one. This may, at least partially,
explain why the ATLAS collaboration obtained small $\sigma_{eff}$
parameter when omitting the higher-order two-gluon exchange mechanism
and feed down for double $\chi_c$ production.
Clearly more studies are needed.

\section{Conclusions}

In this talk a discussion on mechanisms relevant for double $J/\psi$ 
production was presented. Several mechanisms were discussed.
In addition to the commonly recognized leading-order (box) contribution 
and discussed recently by different authors double-scattering mechanism
we included also two-gluon exchange mechanism and a feed down
from double $\chi_c$ channel. The two-gluon exchange mechanism,
formally three loop type, was calculated in the high-energy approximation.
It turned out to be important at high rapidity distance between
two $J/\psi$ mesons.
The two-$\chi_c$ production was considered recently only by our group.
The amplitudes for off-shell gluons $g^* g^* \to \chi_c(J_i) \chi_c(J_j)$
were calculated using $g^* g^* \to \chi_c(J_k)$ vertices calculated
within nonrelativistic pQCD approach. The details of the matrix element
will be shown elsewhere \cite{CSS2017}.
The $t$- or $u$-channel gluon exchange mechanisms lead to a population of
two $\chi_c$ mesons at large rapidity distance.

Both two-gluon exchange and the double feed down mechanisms have
characteristics similar to the double-scattering mechanism.
This may potentially explain why fits to experimental data,
where only leading-order (box) and double scattering mechanisms
were included, found very small value of $\sigma_{\rm eff}$.
Our preliminary calculation strongly suggests that the missing single
parton scattering mechanisms should be included in trials to understand
experimental data. This will be discussed elsewhere \cite{CSSB2018}.

\Acknowledgements 

The work presented here was partially supported by the Polish National
Science Centre grant DEC-2014/15/B/ST2/02528 and the Center for
Innovation and Transfer of Natural Sciences and Engineering Knowledge in
Rzesz\'ow.

\end{document}

%% file: econfmacros.tex
\def\beq{\begin{equation}}
\def\eeq#1{\label{#1}\end{equation}}
\def\eeqn{\end{equation}}

\def\beqa{\begin{eqnarray}}
\def\eeqa#1{\label{#1}\end{eqnarray}}
\def\eeqan{\end{eqnarray}}

\let\bar=\overbar

\def\Dslash{\not{\hbox{\kern-4pt $D$}}}
\def\dslash{\not{\hbox{\kern-2pt $\del$}}}

\def\msb{{\bar{\ssstyle M \kern -1pt S}}}




%% file: eprint_szczurek.bbl
\begin{thebibliography}{99}
\bibitem{review}
N. Brambilla et al.,
Eur. Phys. J. {\bf C71} (2011) 1534.

\bibitem{D0_jpsijpsi}
V.M. Abazov et al. (D0 Collaboration),
Phys. Rev. {\bf D90} (2014) 111101(R).

\bibitem{LHCb_jpsijpsi_7TeV}
R. Aaij et al. (LHCb Collaboration),
Phys. Lett. {\bf B707} (2012) 52.

\bibitem{CMS_jpsijpsi}
V. Khachatryan et al. (CMS Collaboration),
JHEP {\bf 1409} (2014) 094.

\bibitem{ATLAS_jpsijpsi}
M. Aaboud et al. (ATLAS Collaboration), 
Eur. Phys. {\bf C77} (2017) 76.

\bibitem{LHCb_jpsijpsi_13TeV}
R. Aaij et al. (LHCb Collaboration),
arXiv:1612.074451 [hep-ex].

\bibitem{BSZ2011}
S.P. Baranov, A.M. Snigirev and N.P. Zotov,
Phys. Lett. {\bf B705} (2011) 116.

\bibitem{Astalos}
R. Astalos et al..
Proceedings of the Sixth International Workshop on Multiple
Partonic Interactions at the Large Hadron Collider'',
arXiv:1506.05829 [hep-ph].

\bibitem{BSZSS2013}
S.P. Baranov, A.M. Snigirev, N.P. Zotov, A. Szczurek and W. Sch\"afer,
Phys. Rev. {\bf D87} (2013) 034035.

\bibitem{JJ_kt}
S.P. Baranov, Phys. Rev. {\bf D84} (2011) 054012.

\bibitem{KKS2011}
C.H. Kom, A. Kulesza and W.J. Stirling,
Phys. Rev. Lett. {\bf 107} (2011) 082002.

\bibitem{BK2016}
C. Borschensky and A. Kulesza,
Phys. Rev. {\bf D95} (2017) 034029.

\bibitem{C2017}
A. Cisek, a talk at this conference.

\bibitem{CSS2017}
A. Cisek, W. Sch\"afer and A. Szczurek,
a paper in preparation.

\bibitem{PDG}
K.A. Olive et al. (Particle Data Group),
Chin. Phys. {\bf C38} (2014) 090001.

\bibitem{CSSB2018}
A. Cisek, W. Sch\"afer, A. Szczurek and S. Baranov,
a paper in preparation.

\end{thebibliography}
